\documentclass[conference]{IEEEtran}
\IEEEoverridecommandlockouts
\usepackage{cite}
\usepackage{amsmath,amssymb,amsfonts}
\usepackage{algorithmic}
\usepackage{graphicx}
\usepackage{textcomp}
\usepackage{xcolor}
\usepackage{booktabs}
\usepackage{lipsum}
\usepackage[normalem]{ulem}

\def\BibTeX{{\rm B\kern-.05em{\sc i\kern-.025em b}\kern-.08em
    T\kern-.1667em\lower.7ex\hbox{E}\kern-.125emX}}

\usepackage{comment}
    
\begin{document}

\title{Early Assessment of Artificial Lower Extremity Sensory Response Times and Proprioceptive Acuity via Sensory Cortex Electrical Stimulation\\
\thanks{This work was funded by the National Science Foundation (Award No. 1646275)}
}

\author{
\IEEEauthorblockN{ Won Joon Sohn\IEEEauthorrefmark{1}\IEEEauthorrefmark{6}, Jeffrey Lim\IEEEauthorrefmark{1}, Po T. Wang\IEEEauthorrefmark{1}, Susan J. Shaw\IEEEauthorrefmark{3}, Michelle Armacost\IEEEauthorrefmark{3}, Hui Gong\IEEEauthorrefmark{3}, Brian Lee\IEEEauthorrefmark{3}\IEEEauthorrefmark{4}, \\ Darrin Lee\IEEEauthorrefmark{3}\IEEEauthorrefmark{4}, 
Payam Heydari\IEEEauthorrefmark{1}, Richard A. Andersen\IEEEauthorrefmark{5}, Charles Y. Liu\IEEEauthorrefmark{3}\IEEEauthorrefmark{4}, Zoran Nenadic\IEEEauthorrefmark{1}, and An H. Do\IEEEauthorrefmark{2}}
\IEEEauthorblockA{\IEEEauthorrefmark{1}University of California -- Irvine (UCI), Irvine, CA, USA\\
}
\IEEEauthorblockA{\IEEEauthorrefmark{6}Work done at UCI, currently at Abbott Laboratories, Plano, TX, USA}
\IEEEauthorblockA{\IEEEauthorrefmark{2}UCI School of Medicine, Irvine, CA, USA\\ 
Email: and@uci.edu
}
\IEEEauthorblockA{\IEEEauthorrefmark{3}Rancho Los Amigos National Rehabilitation Center, Downey, CA, USA
}
\IEEEauthorblockA{\IEEEauthorrefmark{4}Keck School of Medicine of University of Southern California (USC), USC Neurorestoration Center, Los Angeles, CA, USA}
\IEEEauthorblockA{\IEEEauthorrefmark{5}California Institute of Technology, Pasadena, CA, USA
}

} 

\maketitle

\begin{abstract}
Bi-directional brain computer interfaces (BD-BCIs) may restore brain-controlled walking and artificial leg sensation after spinal cord injury. Current BD-BCIs provide only simplistic "tingling" feedback, which lacks proprioceptive information to perceive critical gait events (leg swing, double support). This information must also be perceived adequately fast to facilitate timely motor responses. Here, we investigated utilizing primary sensory cortex (S1) direct cortical electrical stimulation (DCES) to deliver leg proprioceptive information and measured response times to artificial leg sensations.
Subjects with subdural electrocorticogram electrodes over S1 leg areas participated in two tasks: (1) Proprioceptive acuity: subjects identified the difference between DCES-induced percepts emulating various leg swing speeds; (2) Sensory response: measuring subjects' response time to DCES-induced leg sensations, with DCES-hand, visual and auditory control conditions.
Three subjects were recruited. Only one completed the proprioceptive assessment, achieving 80\%, 70\%, 60\%, and 53\% accuracy in discriminating between fast/slow, fast/medium, medium/slow, and same speeds, respectively (p-value=1.9$\times$10$^{-5}$). Response times for leg/hand percepts were 1007$\pm$413/599$\pm$171 ms, visual leg/hand responses were 528$\pm$137/384$\pm$84 ms, and auditory leg/hand responses were 393$\pm$106/352$\pm$93 ms, respectively.
These results suggest proprioceptive information can be delivered artificially, but perception may be significantly delayed. Future work should address improving acuity, reducing response times, and expanding sensory modalities.

\end{abstract}

\begin{IEEEkeywords}
bi-directional brain-computer interface, electrocorticography, direct cortical electrical stimulation, spinal cord injury, artificial sensation
\end{IEEEkeywords}

\section{Introduction}
\vspace{-0.05in}
Spinal cord injury (SCI) typically causes both a loss of motor and sensory function in the lower extremities.  Although the motor aspect of brain-controlled walking after SCI has received significant attention, the sensory aspect has not. However, the restoration of gait motor function without the simultaneous restoration of sensory function is highly suboptimal, as sensory input plays a critical role in effective and adaptive gait. 
Early evidence demonstrated that bidirectional brain-computer interfaces (BD-BCI) may restore both motor and sensory functions \cite{flesher2021brain,lim2024early,limj:25}. However, the current implementation of artificial sensation in gait function is highly rudimentary, typically only providing a non-specific ``tingling'' or ``buzzing'' sensation in the lower extremities \cite{lim2024early,limj:25} via direct cortical electrical stimulation (DCES) of the primary sensory cortex (S1). Normal gait function is highly dependent on detecting detailed sensory information such as leg position and critical gait events such as double support. However, current BD-BCI technology does not convey such information and is therefore highly suboptimal. Furthermore, given that perturbations in gait may require rapid motor responses, artificial sensation must be delivered and perceived with adequate speed. Almost nothing is known about the response time durations for lower extremity artificial sensation. 
The closest information is prior work from \cite{caldwelldj:19} demonstrating that DCES of hand sensory areas using electrocorticography results in slow response times.

To address these shortcomings and knowledge gaps, this study evaluated whether it is possible to deliver artificial lower extremity sensation containing proprioceptive information. In addition, we also examined the response times to leg artificial sensation. To this end, we utilized the programmable sensory stimulation module from a previously custom-built embedded BD-BCI system \cite{sohn_benchtop_2022} to deliver modulated DCES of S1 to mimic the sensation underlying different velocities of leg swing. S1 DCES was also used to elicit artificial leg sensation while response time was measured and compared to that of auditory and visual stimulation.

\section{Methods}\label{sec:methods}
\subsection{Overview} \label{sec:system}
\vspace{-0.05in}
Subjects undergoing electrocorticogram (ECoG) implantation for epilepsy surgery evaluation over the S1 area were recruited for this study. First, in a proprioceptive acuity task, representations of leg swing at 3 different maximum speeds were converted into stimulation pulse trains using a rate encoding scheme. 
Pairs of stimulation patterns corresponding to different leg swing speeds were delivered and subjects were asked to identify the relative difference between the sensations. Response accuracy was used to assess the proprioceptive acuity. In a second sensory task, subjects' response time to artificial leg sensation was assessed. Here, subjects were asked to react with a button press as soon as they perceived 
 the artificial sensation. The response time was measured over multiple trials and compared to response times for control stimuli, including visual and auditory stimulation.

\subsection{Hardware Design} \label{sec:hardware}
\vspace{-0.05in}
 The BD-BCI was implemented on a custom-designed PCB as in \cite{sohn_benchtop_2022,lim2024early}). Briefly, onboard microcontrollers (48 MHz, Microchip, Chandler, AZ) executed all system functions - most relevant of which is the DCES module. 
 This  module was realized as a combination of charge pump cascade, programmable current source, and programmable H-bridge.
 Wireless transceivers (HOPE Microelectronics, Xili, ShenZhen, China) enabled communication with a base station computer. A custom GUI was used to set all stimulation parameters and deliver stimulation. More specifically, experimenters select the bipolar stimulation channel (any  electrode pair) and waveform parameters (frequency, anodic/cathodic pulse width, current amplitude, pulse train duration). The BD-BCI stimulator delivers biphasic square-pulse trains according to the selected parameters. 

\subsection{Subjects} \label{sec:decode}
\vspace{-0.05in}
Subjects were recruited from a patient population undergoing ECoG implantation for epilepsy surgery evaluation with electrode coverage of the S1 area with expected leg representation.

\subsection{Sensory Stimulation Mapping Procedure}\label{sec:stim}
\vspace{-0.05in}
Clinical cortical stimulation mapping was leveraged to identify stimulation parameters with  sensory responses. This stimulation mapping procedure was performed by clinicians to identify eloquent cortex to inform clinical decision making. The ability to elicit sensory percepts was reaffirmed using the BD-BCI stimulator (device shown to be equivalent to clinical stimulator in \cite{sohn_benchtop_2022}). 
Channels and stimulation parameters which resulted in lower extremity and hand sensory percepts were noted for subsequent verification and use for sensory tasks described further below.
In a typical clinical mapping procedure, adjacent pairs of ECoG electrodes in S1 were sequentially connected to a clinical cortical stimulator (Natus Cortical Stimulator, Natus, Middleton, WI, USA) to form bipolar stimulation channels. For each channel, stimulation was delivered using 1-5 s pulse trains, with systematically varied current amplitudes (0-20 mA), nominal pulse frequency 50 Hz, and nominal pulse width of 250 $\mu$s/phase. 
For each parameter set, subjects were asked for a verbal description of any elicited sensations. Parameter sets eliciting sensation in the lower extremity and hand contralateral to ECoG implantation were noted. Ultimately, one channel and stimulation parameter set that elicited robust sensation was chosen for the sensory tasks below. 

\subsection{Encoding leg swing velocity into artificial sensation}
\vspace{-0.05in}
To objectively assess the ability to deliver artificial proprioceptive information pertaining to leg swing velocity, pre-recorded stimulation pulse trains based on representations of different leg swing velocity were first generated. 
Note that although proprioception was not tested explicitly (e.g., through joint angle identification), the perception of limb movement speeds is heavily based on proprioception in able-bodied individuals \cite{kerrgk:02,horvatha:23}.
Three representations were created to simulate fast, medium, and slow leg swings. Each representation consists of a ``bell-shaped" time series of similar duration (1.9 s), but with differing peak amplitudes to represent different maximum leg swing velocities. These bell-shaped profiles were then converted into a stimulation pulse train by using an Izhikevich spiking neuron model \cite{izhikevichem:03}. The Izhikevich neuron model was selected due to its balance between biological realism and computational efficiency, making it well-suited for implementation on our embedded BD-BCI hardware \cite{sohn_benchtop_2022}. Furthermore, this approach is consistent with the rate encoding scheme typically seen in sensory systems \cite{lieberjd:20}. 

\subsection{Proprioceptive Acuity Task}\label{sec:proprioception}
\vspace{-0.05in}
To assess the ability to deliver proprioceptive information via artificial sensation, we utilized a leg swing velocity discrimination task.
The stimulation channel and parameters previously identified to elicit artificial leg percepts were used to deliver pairs of stimulation trains derived from the fast, medium, and slow velocity representations.
Note that there are 4 possible combinations: fast vs medium, fast vs slow, medium vs slow, and identical speeds (the order in which the two stimuli were delivered in each pair was randomized). 
There is an 50\%, 200\%, and 100\% difference in maximum velocity between medium and fast, slow and fast, and slow and medium speeds, respectively.
After each stimulation pair, the subject was asked to identify whether the second stimulation felt faster, slower, or the same as the first. The subjects' ability to discriminate  between different leg swing speeds (proprioceptive acuity) was determined based on the accuracy of response. 
For each 4 stimulus combination, at least 10 trials were administered, depending on subject availability. 
The statistical significance of the subjects' response was assessed empirically by comparing their performances to that of 10$^6$ random guess trials.

\subsection{Artificial Sensory Response Time Task}
\vspace{-0.05in}
To assess the sensory response time, subjects were presented with 3 sensory modalities across 2 locations. The modalities included artificial sensation by S1 DCES, visual stimulation, and auditory stimulation. The locations included leg and hand. Leg artificial sensation is the primary assessment, while the remaining conditions served as controls. The stimulation pulse train had a fixed duration of 500 ms. For
leg/hand DCES, subjects were asked to press a foot switch/keyboard button as quickly as possible after perceiving the artificial sensation of the leg/hand, respectively.  
Response times for visual and auditory modalities for the leg/hand locations were also established in a similar manner. 
All response times were measured as the time duration from onset of stimulus to the onset of either the foot switch response (subject presses foot switch by plantarflexing foot while in semi-recumbent position in hospital bed for leg location conditions) or keyboard press (subject presses a designated keyboard button). For the auditory condition, the stimulus was a 1 kHz monotonous tone with a fixed duration of 100 ms. For the visual condition, the stimulus consisted of the screen color changing from black to green. 
In the DCES modalities, stimuli were delivered using the BD-BCI GUI on the base station computer with the experimenter using a silent click mouse to mitigate the effect of subject anticipation. 
Since delivery of auditory and visual modalities was fully automated, the stimulus onset was randomized to mitigate the effect of anticipation or cue timing prediction. Specifically, the stimuli were delivered with random delay between 1 to 4 seconds upon initiation of each trial. 
Each condition was repeated for at least 50 trials, or as tolerated by the subjects. 
The response times were averaged across all trials and compared across conditions (t-test).

\section{Results}

\begin{figure}[!htpb]
    \centering
    \includegraphics[width=\linewidth]{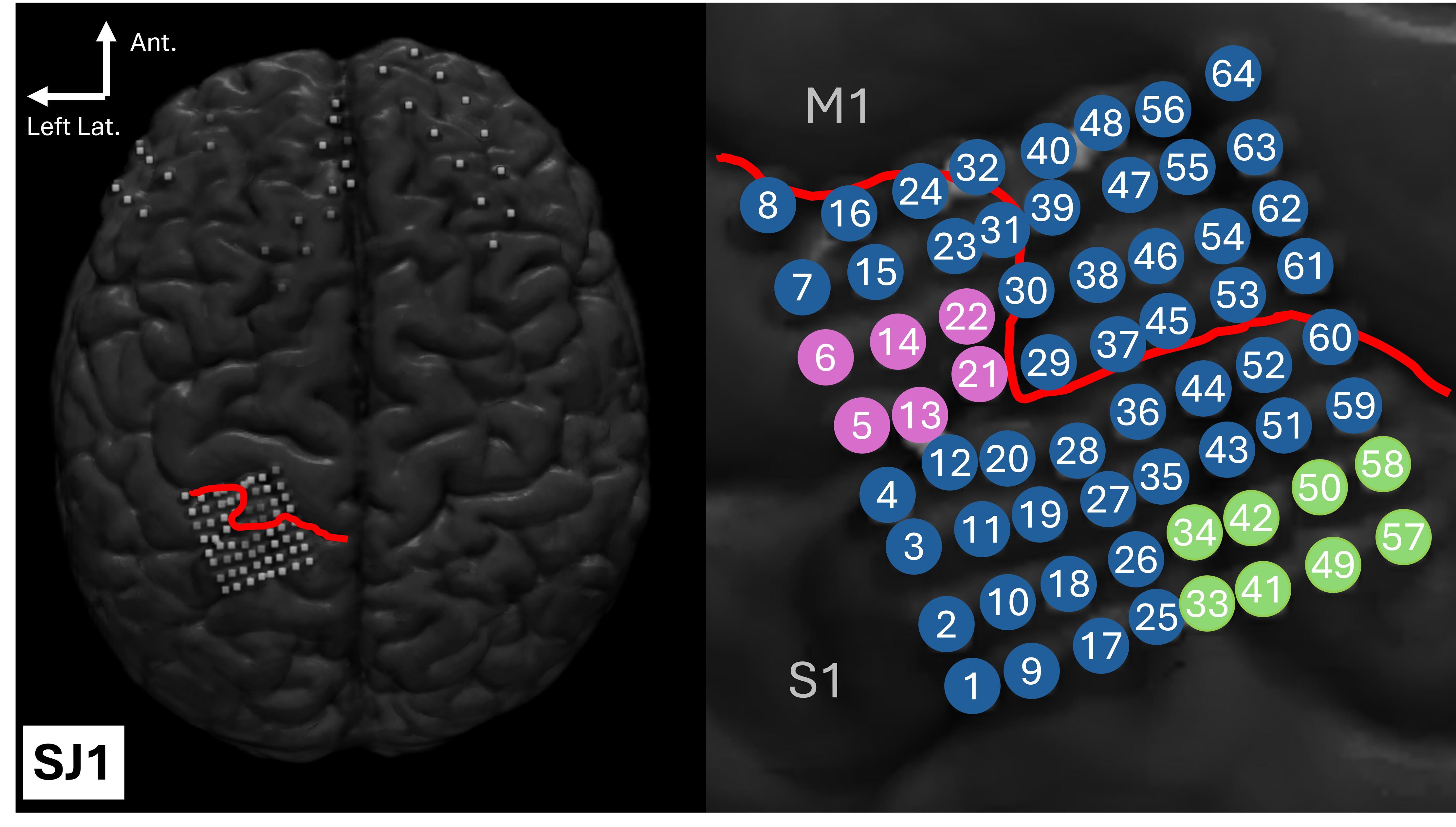}
    \includegraphics[width=\linewidth]{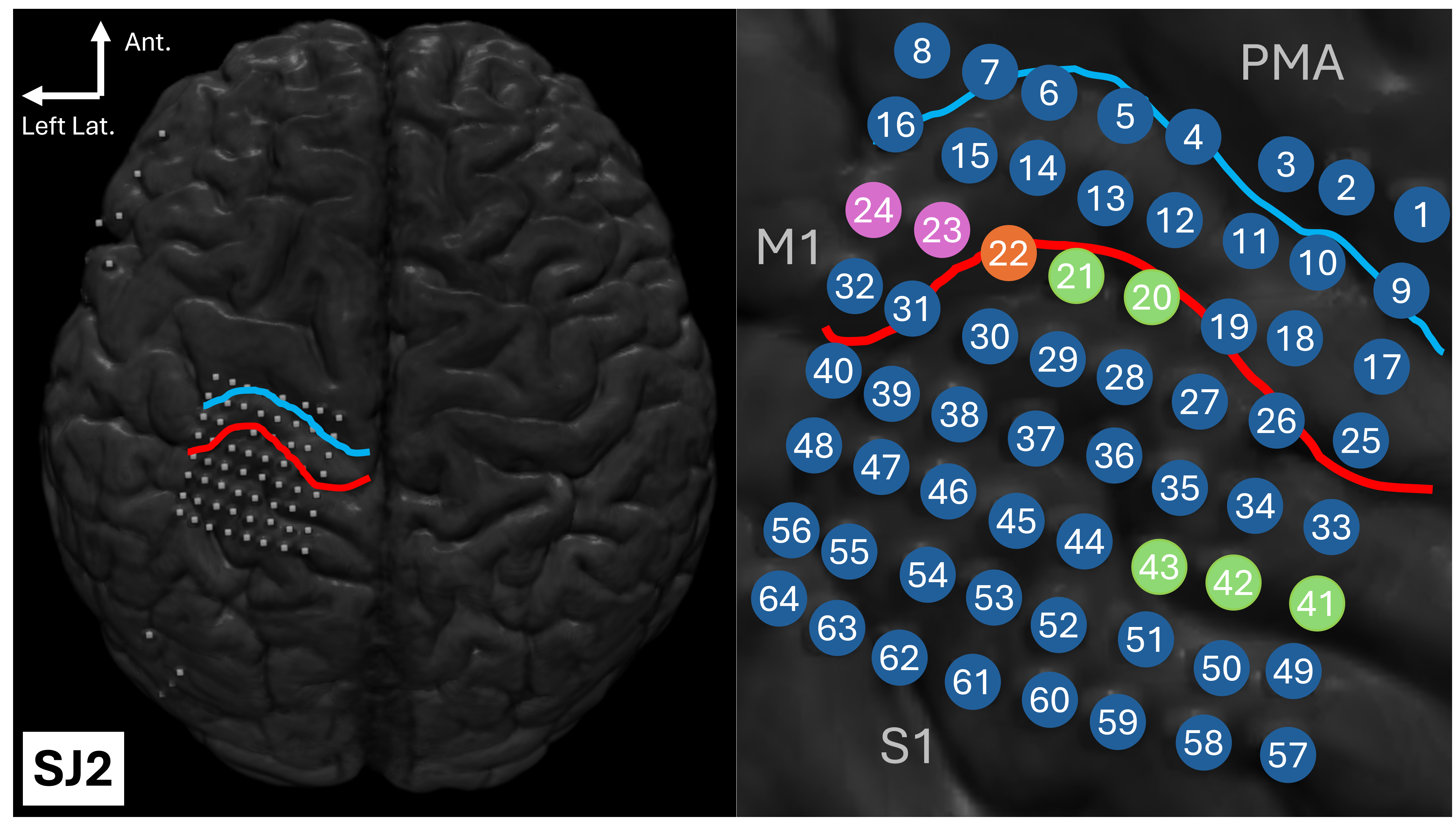}
    \includegraphics[width=\linewidth]{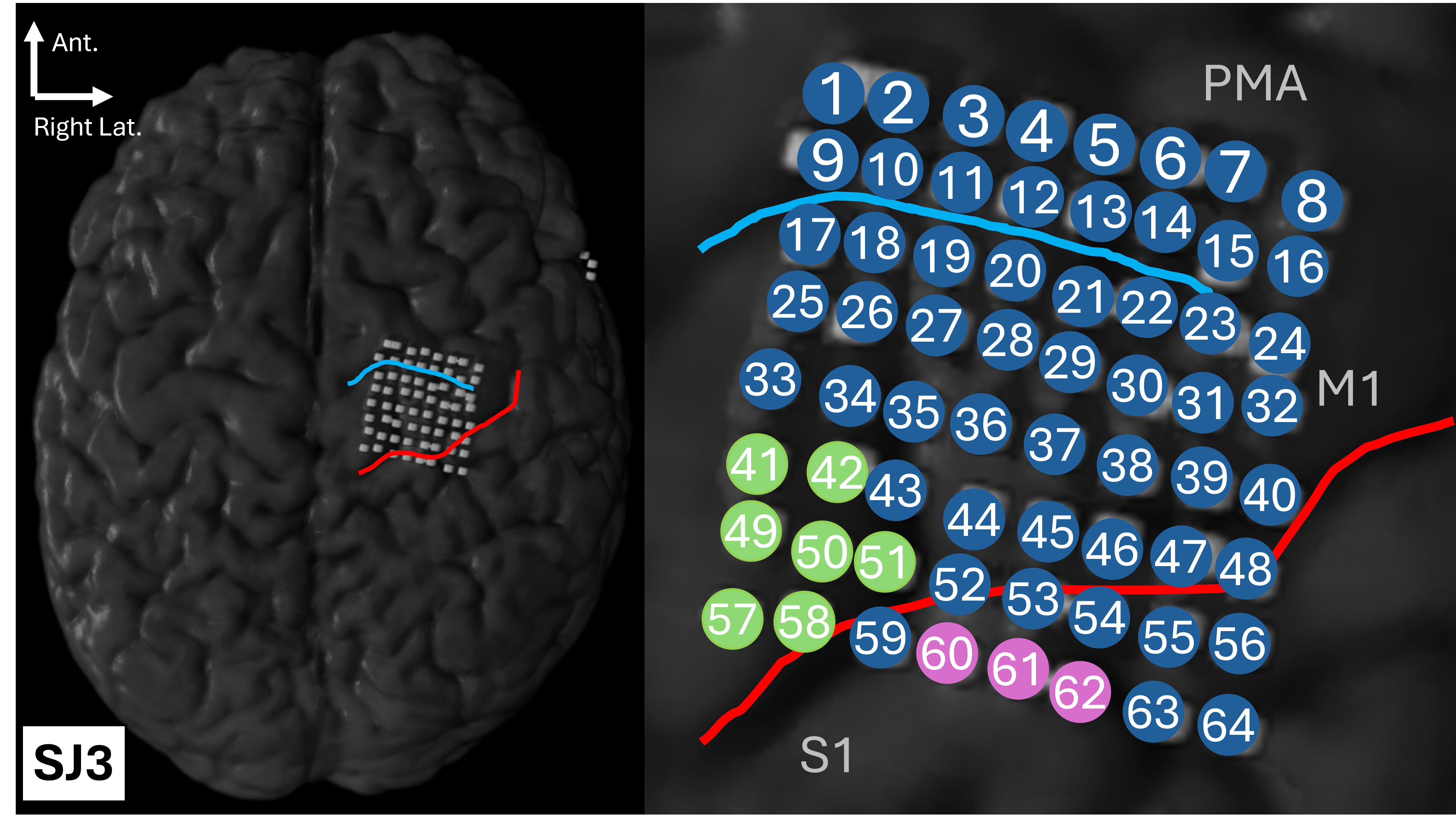}
    \vspace{-0.3in}
    \caption{ECoG electrode locations within each subjects' brain (axial view). Images derived from post-implant CT image and pre-implant MR image. Approximate locations of the pre-central sulcus (light blue) and central sulcus (red) delineate the M1 and S1 cortices. (Green): Electrodes identified to have leg sensory representation. 
    (Purple): Electrodes identified to have hand sensory representation.
    (Orange): Electrodes with both hand and leg sensory representation.
    Additional percept and stimulation parameters in Table \ref{tab:stim_map}. M1: Primary motor cortex. S1: Primary sensory cortex. PMA: Pre-motor area.} 
    \label{fig:coreg}
    \vspace{-0.2in}
\end{figure}

\subsection{ECoG Subject Details}
\vspace{-0.05in}
This study was approved by the IRB of the University of California, Irvine and the Rancho Los Amigos National Rehabilitation Center. A total of 3 subjects undergoing epilepsy surgery evaluation with ECoG implanted over the S1 areas with expected leg representation  were recruited and provided informed consent to participate in the study
(Fig.~\ref{fig:coreg}).
Due to limited subject availability, only 1 subject (SJ1) participated in the proprioceptive acuity task.
For the artificial sensory response time assessment, 3 subjects (including SJ1) participated, but with limited completion across test conditions. 
Subject characteristics and what tasks each subject underwent are summarized in Table \ref{tab:subjects}.

\begin{table}[!htbp]
    \centering
    \caption{Subject Characteristics}
    \begin{tabular}{cccc}
    \toprule
        Subjects & Sex & Age & Sensory Tasks Performed \\
        \midrule
        SJ1 & F & 42 & Proprioceptive, Sensory Response Time \\
        & & & (partial completion) \\
        SJ2 & F & 21 & Sensory Response Time \\
        & & & (partial completion) \\
        SJ3 & F & 32 & Sensory Response Time (all conditions) \\ 
        \bottomrule
    \end{tabular}
    \label{tab:subjects}
\end{table}

\begin{table*}
    \centering
    \caption{Examples of Sensory Percepts from Cortical Mapping Procedure}
    \begin{tabular}[width = \linewidth]{cccccc}
    \toprule
         SJ\# & Stimulation Channel & Amplitude (mA) & Pulse Train Frequency (Hz) & Sensation Location  & Reported Sensation \\
    \midrule
SJ1         &57-58 & 5 & 50 & Anterior lower right leg & Tingling \\
         &49-50 & 6.4 & 100 & Anterior lower right leg & Tingling \\
         &41-42 & 8.6 & 150 & Anterior lower right leg & Tingling, brushing \\
         &33-34 & 8 & 250 & Right knee & Tingling \\
         & 5-6 & 4.9 & 100 & Right palm & Tingling \\
         & 13-14 & 4.0 & 50 & Right palm & Tingling \\
         & 21-22 & 4.1 & 150 & Right hand & Poking \\

    \midrule
    SJ2     & 41-43 & 8 & 50 & Entire right leg & Tingling \\
    & 41-42 & 10 & 50 & Right posterior lower leg & Tingling\\
        & 20-21 & 10 & 50 & Right posterior upper leg & Tingling\\
        & 21-22 & 6.8 & 50 & Right posterior upper leg & Tingling \\
        & 22-23 & 6  & 50 & Right palm & Tingling \\
        & 23-24 & 6.8 & 50 & Right palm & Tingling \\
        \midrule
    SJ3 & 41-49 & 1.8 & 50 & Left knee & Pulsing\\
        & 41-42 & 2.2 & 50 & Left knee & Pressure \\

        & 49-50 & 3.4 & 50 & Left lower posterior leg & Tingling \\

        & 50-51 & 3.0 & 50 & Left knee & Pulsing \\

        & 57-58 & 1.6 & 50 & Entire left leg & Brushing \\
         & 59-60 & 1.8 & 50 & Left hand & Brushing \\
         & 61-62 & 0.3 & 50 & Left hand & Brushing \\
        
    \bottomrule
    \end{tabular}
    \label{tab:stim_map}
\end{table*}

\begin{table*}[!htpb]
    \centering
    \caption{Stimulation parameters utilizes for study procedures. C: Channel I: current, PW: pulsewidth, $f$: frequency. *: Leg DCES parameters mistakenly used instead of hand percept during DCES-hand condition for sensory response time study.}
    \label{tab:stimparameters}
    \begin{tabular}{ccccc}
        \toprule
         Subject \# & Task & Parameters & Percept\\         
         \midrule
        SJ1 & Proprioceptive Acuity & C: 49-50, I: 6.2 mA, PW: 250 $\mu$s/phase, $f$: up to 417 Hz & Leg     \\ 
         & Sensory Response Time &  C: 49-50, I: 6.2 mA, PW: 250 $\mu$s/phase, $f$: 50 Hz  & Leg     \\
         & Sensory Response Time &  C: 5-6, I: 4.9 mA, PW: 250 $\mu$s/phase, $f$: 100 Hz  & Hand     \\
         \midrule
        SJ2 & Sensory Response Time &  C: 22-23, I: 6 mA, PW: 250 $\mu$s/phase, $f$: 50 Hz  & Hand      \\
        \midrule
        SJ3 & Sensory Response Time & C: 49-41, I: 1.6 mA, PW: 250 $\mu$s/phase, $f$: 50 Hz   & Leg      \\
            & Sensory Response Time & C: 61-62, I: 0.2 mA, PW: 250 $\mu$s/phase, $f$: 50 Hz   & Hand*     \\
         \bottomrule
    \end{tabular}
\end{table*}

\subsection{Stimulation Mapping Results}
\vspace{-0.075in}
Stimulation mapping was performed across all electrodes of the ECoG grid. Stimulation parameters that elicited hand and leg sensory responses and the subjects' verbal descriptions are reported in Table~\ref{tab:stim_map}, and topographically portrayed in Fig. \ref{fig:coreg}. Note that while the sensory mapping was more extensive, only examples of salient responses were reported here. Also, some motor responses (involuntary movement) were also elicited during the mapping procedure. Channels with motor responses at any stimulation parameter set were excluded from this list. For all subjects, the most common artificial sensation was a tingling sensation.  

\begin{figure}[!htpb]
    \centering
    \includegraphics[width=\linewidth]{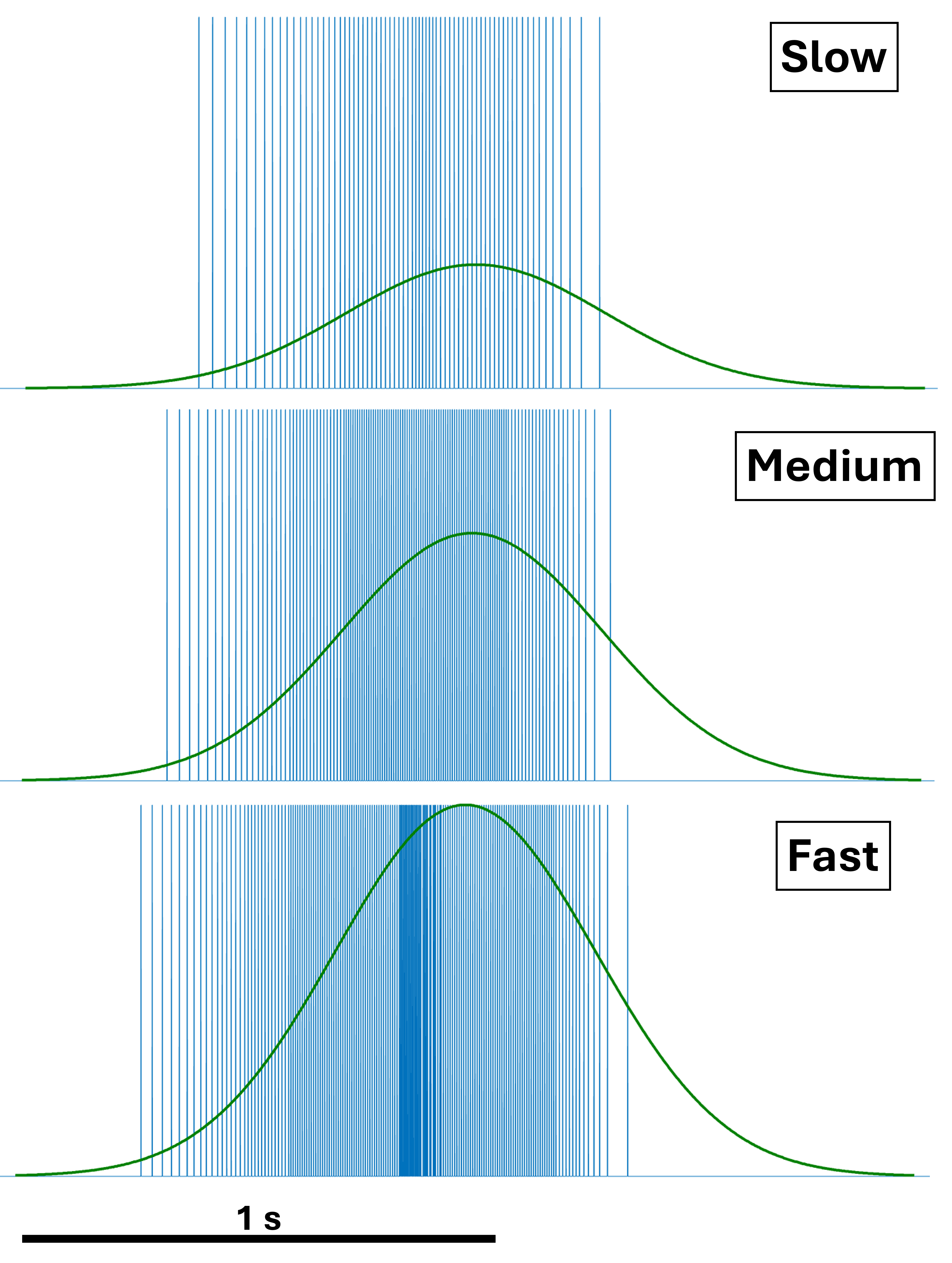}
    \vspace{-0.35in}
    \caption{Representations of slow leg swing velocity time series were translated in stimulation pulse trains. Green: Time series of representation of fast, medium, and slow leg swing speeds. Blue: stimulation pulse train resulting from leg swing representations.}
    \label{fig:IZN}
    \vspace{-0.2in}
\end{figure}

 \subsection{Artificial Proprioceptive Sensory Discrimination}
 \vspace{-0.05in}
The time series for the 3 leg swing speeds were translated into spike trains as in the methods section. An example of the spike train for each velocity level is shown in Fig. \ref{fig:IZN}. 
The resulting pulse trains had the following maximum stimulation frequencies: fast speed up to $\sim$417 Hz, medium speed up to $\sim$213 Hz, and slow speed up to $\sim$141 Hz.
There was $\sim$196\%, $\sim$96\%, and $\sim$51\% difference between maximum stimulation rate between slow and fast, medium and fast, and  slow and medium speeds, respectively. 
Due to limited subject access, only SJ1 could participate in this sensory task.
For SJ1, the stimulation parameters chosen for this task are summarized in Table \ref{tab:stimparameters}. 
A total of 45 sensory discrimination trials were performed and the accuracy of the subject's responses are summarized in Table \ref{tab:accuracy} below. The empiric p-value was $p=1.9\times10^{-5}$.

It was observed that the coarsest (highest) velocity difference, i.e., fast vs slow, resulted in the most accurate discrimination (80\%), while the subject could only correctly discriminate 53\% of trials where the underlying velocity comparison was the same. Also, there was a direct relationship between the difference in the maximum stimulation frequency and the discrimination accuracy.

\begin{table}[!htpb]
    \centering
    \caption{Accuracy of proprioceptive sensory discrimination. SJ1 only.}
    \label{tab:accuracy}
    \begin{tabular}{cccccc}
        \toprule
          & Fast/slow & Fast/medium & Medium/Slow & Same    \\         
         \midrule
         Accuracy & 80\% & 70\% & 60\% & 53\%     \\ 
         \# of Trials & 10 & 10 & 10 & 15  \\
         \bottomrule
    \end{tabular}
\end{table}

\subsection{Artificial Sensory Response Time}
\vspace{-0.05in}
All 3 subjects underwent the sensory response time task. Due to limited subject access or fatigue, some subjects could not perform all sensory response conditions. Also, due to an error in implementing the experimental protocol with SJ3, the stimulation parameters to elicit a leg percept was utilized while testing DCES hand condition. Hence, SJ3's DCES hand sensory response results were excluded from the analysis below. 
The stimulation parameters chosen for each subject are summarized in Table~\ref{tab:stimparameters}.

SJ1 performed 57 DCES-leg, 62 DCES-hand, 17 visual-leg, 50 visual-hand, 53 auditory-hand, and 0 auditory-leg  trials.
SJ2 completed 0 DCES-leg, 46 DCES-hand, 0 visual-leg, 56 visual-hand, 0 auditory-leg, and 55 auditory-hand trials.
SJ3 completed 51 DCES-leg, 53 DCES-hand (excluded from analysis), 53 visual-leg, 50  visual-hand, 49 auditory-foot, and 50 auditory-hand  trials.

\begin{figure}[!htpb]
    \centering
    \includegraphics[width=\linewidth]{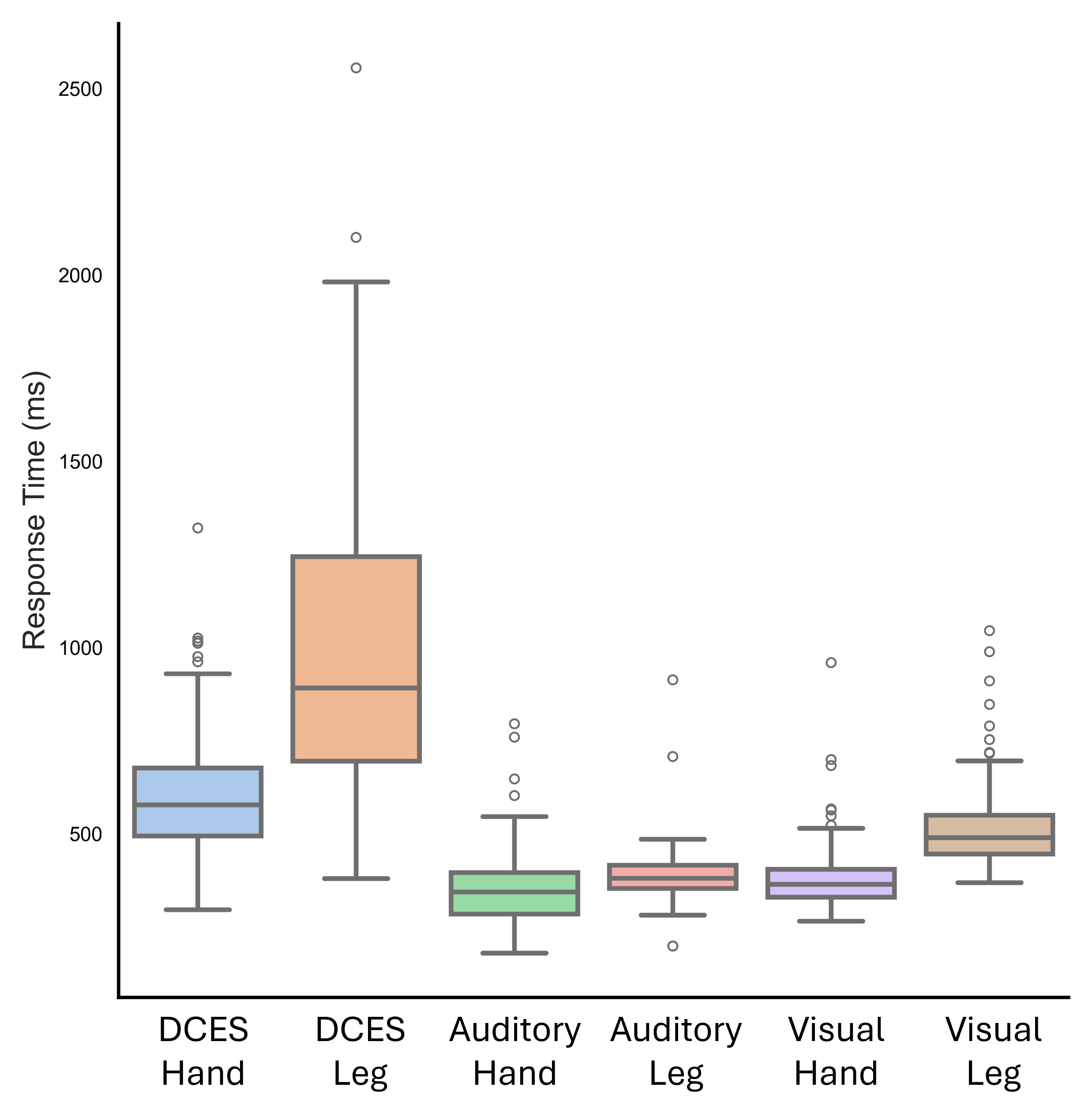}
    \vspace{-0.35in}
    \caption{Average response times for each stimulation modality and location combination. All pairwise comparisons (t-test) were significant an $\alpha$ threshold of 0.01, except for between auditory-leg and visual-hand response times (p-value=0.5).}
    \label{fig:responsetimes}
    \vspace{-0.3in}
\end{figure}

The average artificial sensation response time for the leg was 1007$\pm$413 ms, and 599$\pm$171 ms for the hand.
The average auditory response time for the leg was 393$\pm$106 ms, and 352$\pm$93 ms for the hand. The average visual response time for the leg was 528$\pm$137 ms, and 384$\pm$84 ms for the hand. The average response time across all conditions (excluding SJ3's DCES-hand condition) are shown in Fig. \ref{fig:responsetimes}. All pairwise comparisons (t-test) were significant at an $\alpha$ threshold of 0.01, except for between auditory-leg and visual-hand conditions (p-value=0.5).
Even though SJ3's DCES hand condition was implemented erroneously, we report that the DCES-leg and DCES-hand response times were 1301$\pm$404 and 1034$\pm$409 ms, respectively. 

\section{Discussion}

This study represents the first demonstration of the feasibility of encoding proprioceptive information in artificial leg sensation.   
Specifically, we established that it may be  possible to convey information pertaining to different leg swing speeds through modulated stimulation of the primary sensory cortex (S1). However, we also observed that the perception of artificial sensation can be significantly delayed relative to the onset of simulation. 

Although these findings indicate early feasibility of delivering proprioceptive information to the subject, there is still a need for improvement in the discrimination accuracy.
Notably, subjects demonstrated the highest accuracy when distinguishing between the most disparate gait velocities (i.e., fast vs. slow), but performance declined when differentiating between identical speeds.
Such a finding is consistent with prior literature \cite{johnsonko:81,papg:99,marinif:17} in  sensory discrimination and resolution, where perceptual accuracy improves as stimulus differences increase.
The direct relationship between the difference in maximum stimulation frequency being compared and the subject's discrimination accuracy suggests that rate encoding may be a viable strategy for conveying proprioceptive information. This is consistent with the rate encoding scheme typically seen in S1 \cite{depeaulta:13,lieberjd:20}.
Also, a similar concept was shown in artificial upper extremity sensation \cite{hiremath_human_2017}, although no objective tests were used to quantify sensory discrimination.
Nevertheless, there is still a substantial error in the discrimination between leg swing velocities. Several factors may account for this performance limitation. One cause for the low discrimination accuracy may be due to lack of familiarity to artificial sensation, as it is a novel experience for the recipient. Furthermore, the errors in discrimination may also be due to a lack of adequate timing cues. It is known that the addition of timing cues can help to improve proprioceptive discrimination  \cite{kerrgk:02}. Conversely, the pulse train duration is disconcordant to typical expectations. Specifically, whereas able-bodied individuals would typically expect a longer step duration with slower leg velocities, the subject experienced the opposite here (Fig. \ref{fig:IZN}). Based on these findings and considerations, future work in this area may need to involve sensory training or familiarization time, and the leg trajectories used to generate the pulse trains may need to be derived from motion capture of real human leg trajectories during walking or modeled in a more realistic manner.

The observation that artificial response time were significantly longer than that of both visual and auditory stimulation raises concerns that these delays may lead to downstream delays in motor responses in BD-BCI operation. 
Although not compared directly to somatic sensory response times, historical data from \cite{chapwouot:18} indicate that these times far exceed typical response times by nearly two- or three-fold.  
Similar delayed response times were observed with artificial sensation for the upper extremities in prior work \cite{caldwelldj:19}.
It is possible that the sedating effect of the analgesic and anti-seizure medications that the subjects were on may contribute to the prolonged response times.
Another contributor to response time delay may be suboptimal electrical stimulation power, leading to slow accumulation of the charge needed to activate S1 neurons. Future work will need to identify means to reduce this perception delay by optimizing stimulation parameters and incorporating sensory training paradigms \cite{taylorsc:21}.

The stimulation mapping procedure demonstrated that the leg sensory representation area was typically medial to the hand representation area. Although leg representation is classically thought to be primarily within the interhemispheric S1 region, we found that there is still leg sensory representation lateral to the interhemispheric fissure. Although some sensory representation was found in M1, it should be noted that the accuracy of electrode position identification may be limited when using pre-implantation MRI scans. Furthermore, it has also been well established that sensory representations can be found in putative M1 areas \cite{penfieldw:37}. Although tingling is the most common percept reported during the sensory mapping procedure (Table \ref{tab:stim_map}), this percept alone is not highly informative for sensorimotor processes. As such, modulation using rate encoding, as shown above can leverage this percept to encode useful information.

Several limitations must be acknowledged. First, the study only had a small sample size, wherein the proprioceptive acuity task was conducted with only one subject, and the response times only involved 3 subjects. Second, the stimulation pulse trains were based on synthetic representations rather than real leg swing trajectories obtained from human motion. Third, while artificial sensory response times were compared to visual and auditory stimulation response times, it has not yet been compared to tactile response times, which would offer a more relevant baseline. Finally, this study only examined the possibility of delivering lower extremity proprioception. 

Future work will need to systematically confirm the above findings in a larger sample size, improve on sensory discrimination accuracy and response times, and expand to additional sensory modalities (particularly graded tactility/pressure sensation as such modalities are also critical for normal gait function).

In conclusion, this study demonstrates the feasibility of artificially conveying proprioceptive information to the lower extremities via DCES. However, the significant sensory delays observed indicate the need for further characterization and optimization.  Reducing perception latency is critical to more closely mimic the natural sensorimotor loop and to enable timely motor responses to environmental perturbations.
Future work will seek to use real trajectory information from leg swings rather than simulated representations. In addition, this will also need to be tested a larger population and in the target SCI population with paraplegia and loss of sensation in the lower extremities.  

\bibliographystyle{ieeetr}
\bibliography{perceptsArXiv}

\end{document}